\documentclass[conference]{IEEEtran}
\IEEEoverridecommandlockouts
\usepackage{cite}
\usepackage{amsmath,amssymb,amsfonts}
\usepackage{algorithmic}
\usepackage{graphicx}
\usepackage{textcomp}
\usepackage{xcolor}
\usepackage{booktabs}   
\usepackage{tabularx}
\usepackage{subcaption}
\usepackage{enumitem}

\def\BibTeX{{\rm B\kern-.05em{\sc i\kern-.025em b}\kern-.08em
    T\kern-.1667em\lower.7ex\hbox{E}\kern-.125emX}}
\begin{document}

\title{Experimental Validation and Mitigation of RRC Storm Attacks in 5G Cellular Networks}
%From Emulation to Practice: Reproducing and Mitigating 5G RRC Signaling Storms
%Revisiting 5G RRC Signaling Storms:\\Practical Reproduction and Mitigation Implementation
% \author{
%     \IEEEauthorblockN{Abdallah Abou Hasna, Ammar El Falou
% \IEEEauthorblockA{CEMSE Division, Kinhttps://www.overleaf.com/project/69e32b4a1efae0ac96c12799#g Abdullah University of Science and Technology (KAUST), Saudi Arabia}}
% Email: \{abdallah.abouhasna, ammar.falou\}@kaust.edu.sa}
\author{
    \IEEEauthorblockN{Abdallah Abou Hasna, Ammar El Falou}
\IEEEauthorblockA{CEMSE Division, King Abdullah University of Science and Technology (KAUST), Saudi Arabia}
Email: \{abdallah.abouhasna, ammar.falou\}@kaust.edu.sa}

\maketitle

\begin{abstract}
The initial access phase of the 5G system remains sensitive because the base station (gNB) must allocate radio resources before the user is fully authenticated. In particular, the random access channel (RACH) procedure can be abused to generate large numbers of incomplete connection attempts, creating a signaling storm that consumes gNB resources and prevents legitimate users from connecting successfully. In this paper, we implement this signaling storm attack using the OpenAirInterface project and validate it on a real testbed composed of software-defined radios and commercial phones. We then design and implement a lightweight mitigation technique that operates directly at the gNB by monitoring and acting on suspicious half-open connections. To make the system observable in practice, we also develop a network management interface that visualizes the network state in real time and highlights suspicious activity during the attack phase. Finally, the work is released as open source so that other researchers can reproduce our results, build on the implementation, and evaluate new mitigation strategies.
\end{abstract}


%\vspace{-0.15cm}
\section{Introduction}
The first contact between a user equipment (UE) and a 5G base station (gNB) occurs at the radio access side, before the UE is fully authenticated by the network~\cite{3gpp-33.501, 3gpp-38.331,abouhasna2026}. At that point, the gNB must process early radio resource control (RRC) signaling and provide an uplink grant to allow the UE to establish a connection. In particular, the random access channel (RACH) procedure is performed first~\cite{3gpp-38.300}, and the resulting uplink transmission carries the initial registration request message to initiate authentication with the core network. This early access path is fundamental, but it is also exposed by design: resources are allocated before the UE is reliably authenticated.

This vulnerability matters because an attacker does not need to break 5G cryptography to affect the network. By repeatedly triggering the RACH procedure, a malicious UE can create many incomplete connections and keep the gNB busy with requests that never finish. If this is done efficiently, the cell starts rejecting legitimate connection requests.

This type of attack is called signaling storm~\cite{tabiban-storm}, and it has already been studied in the literature. In particular, ~\cite{rrcstormdetection} presented the RRC signaling storm attack, implemented and evaluated using the OpenAirInterface (OAI) project~\cite{oai}. They also proposed a detection method based on some access protocol features, deployed within an O-RAN (Open Radio Access Network) architecture as an external application. However, their work focuses solely on detection, and evaluation is conducted in an emulated setup rather than a real radio testbed. In addition, their implementation is not available as open source, which limits reproducibility and further development. 

In this paper, we reproduce the RRC signaling storm attack in OAI and refine its implementation to make it work in realistic conditions. We validate the attack on a real testbed, where additional timing constraints appear, compared with emulation, which requires specific handling in the attacker implementation.
We go beyond detection by designing and implementing a mitigation technique that runs directly at the gNB. Instead of relying on heavy mechanisms or full user identification, the mitigation technique monitors patterns of incomplete connections in incoming requests and uses them to protect the gNB from overload. We evaluate both the attack and the mitigation using software-defined radios (SDRs) and commercial phones, which allows us to observe the system behavior under realistic radio conditions.
\\We also add a network management system (NMS) that provides real-time visibility into the experiment by showing system status during normal operation and relevant information about the attack and mitigation. This makes the whole setup easier to analyze and closer to a real operational scenario. Additionally, our full implementation will be released as open source to support reproducibility and future extensions. 

In summary, this paper makes four main contributions:
\begin{itemize}
    \item We provide an open-source implementation of the RRC signaling storm attack using OAI, including the practical refinements needed to make the attack work in practical scenarios.
    \item We design and implement a mitigation technique that works directly at the gNB during the early access phase.
    \item We validate both the attack and the mitigation technique on a real testbed using SDRs and commercial devices.
    \item We develop an NMS that displays the system state during normal and attack conditions.
\end{itemize}
Together, these contributions move the RRC signaling storm from an emulated problem to a practical, reproducible 5G scenario with a working defense.

%\vspace{-0.1cm}
\section{Background}

\subsection{Initial Access Phase}

\begin{figure}
%\vspace{-0.35in}
    \centering
    \includegraphics[width=1.05\linewidth]{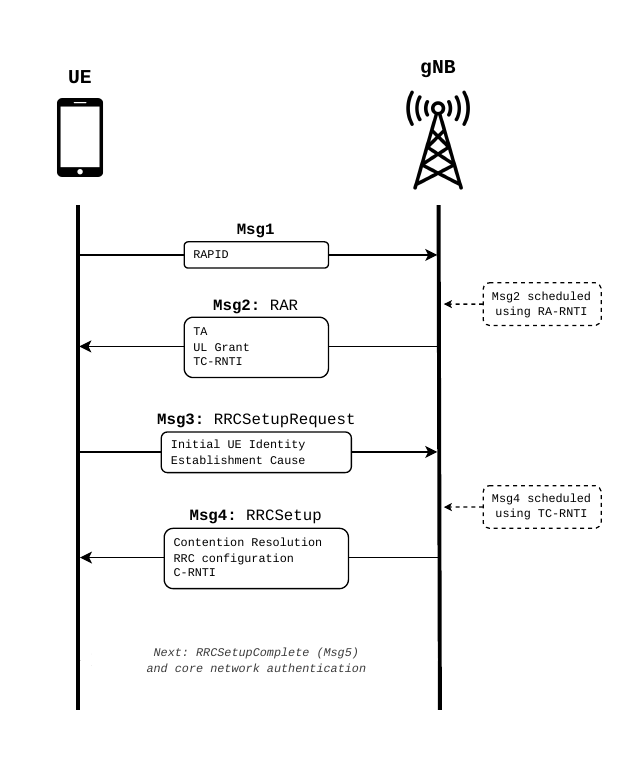}
     \vspace{-1cm}
    \caption{Normal 5G RACH procedure.}
    \label{fig:rach_procedure}
\vspace{-0.2in}
\end{figure}

In 5G, the UE first synchronizes to the cell and reads broadcast information before it can request service. In particular, the system information block type 1 (SIB1) contains essential network information, including common radio configuration and initial access parameters. Once this message has been received, the UE can start the RACH procedure to establish uplink synchronization and obtain initial radio resources~\cite{3gpp-38.331, 3gpp-38.300}. This procedure is composed of four main steps~\cite{3gpp-38.321,3gpp-38.331}. Fig.~\ref{fig:rach_procedure} illustrates this procedure:
\begin{enumerate}
    \item The UE starts with \textbf{Msg1}, which contains a randomly selected preamble identifier (RAPID). Both the available preambles and the time-frequency resources for Msg1 are configured through SIB1.
   \item The gNB responds with \textbf{Msg2}, called the random access response (\textbf{RAR}). This response includes the RAPID that matches the transmitted preamble, a timing advance (TA) command to adjust the UE uplink timing, an uplink grant that allows the UE to transmit the next message, and a temporary cell radio network temporary identifier (\textbf{TC-RNTI}).
    \item Using the allocated resources, the UE sends \textbf{Msg3}, which carries the \textbf{RRCSetupRequest} message. This is the first higher-layer message, and it includes the establishment cause and an initial UE identity, which can be either part of the 5G-S-TMSI (SAE Temporary Mobile Subscriber Identity) or a random value.
    \item Finally, the gNB responds with \textbf{Msg4}, the \textbf{RRCSetup} message. This message performs contention resolution by confirming the identity received in Msg3, since multiple UEs may have selected the same preamble in Msg1. It also carries the initial RRC configuration and assigns the UE a permanent C-RNTI. At this point, the gNB allocates a UE context for this connection attempt. This context refers to the internal memory and reserved resources used to track the UE connection. The connection then moves to the next stage, where authentication with the core network starts through messages such as \textbf{RRCSetupComplete}, often referred to as \textbf{Msg5} in the literature.
\end{enumerate} 

\subsection{Scheduling and Temporary identifiers}

Between the main messages, the actual scheduling is carried by downlink control information (DCI) messages. Each scheduling message is associated with a specific RNTI, which is used to address transmissions.

After Msg1, the gNB schedules Msg2 using a DCI scrambled with a random access RNTI (RA-RNTI), which is derived from the time and frequency resources used for Msg1 transmission. This identifier is not specific to a single UE and may correspond to multiple devices transmitting on the same occasion.
Later in the procedure, the transmission of Msg4 is scheduled using a DCI scrambled with the TC-RNTI assigned in Msg2. After Msg4, subsequent signalling, including Msg5, is scheduled using the assigned C-RNTI.

A summary of the main identifiers and information involved in the RACH procedure is given in Table~\ref{tab:rach_info}.

\begin{table}
\vspace{0.06in}
\centering
\normalsize
\caption{Information available during early access}
\label{tab:rach_info}
\begin{tabular}{|c|p{5.5cm}|}
\hline
\textbf{Step} & \textbf{Information known to gNB} \\
\hline
Msg1 & RAPID, RA-RNTI \\
\hline
Msg2 & TA, TC-RNTI \\
\hline
Msg3 & UE initial identity, establishment cause \\
\hline
\end{tabular}
%\vspace{-0.2cm}
\end{table}

\section{RRC Signaling Storm Attack}

The RRC signaling storm attack exploits the limited information available to the gNB during the early access phase. As described in the previous section, the UE identity carried in Msg3 can be a random value, and the gNB cannot authenticate the UE before the connection establishment continues toward the core network~\cite{3gpp-38.300,3gpp-33.501}. Therefore, a malicious UE can repeatedly start new RACH procedures, each time appearing as a new connection attempt.

In a normal connection, the UE sends Msg5 after receiving Msg4. In the attack case, the malicious UE intentionally stops before this step and restarts the random access procedure. As a result, the gNB allocates resources and creates a UE context for a connection that never completes. We refer to this state as a half-open connection.
\\If the attacker repeats this process, the number of half-open connections grows until the gNB reaches its supported UE capacity. At that point, new legitimate UEs may be rejected because the gNB cannot allocate additional UE contexts.

The prior work assumes that after sending Msg4, the gNB keeps the UE context for a waiting time while expecting Msg5~\cite{rrcstormdetection}. However, the 3GPP specifications do not define a dedicated timer for this exact waiting period. Therefore, in this paper, we treat this timer as an implementation-dependent behavior because, in practice, a gNB still needs some cleanup logic for incomplete connections. Otherwise, a UE that never completes the connection would leave resources reserved indefinitely.

Under this assumption, the attacker must create enough half-open connections before the first reserved context is released. Otherwise, the process becomes ineffective: while the attacker creates new half-open connections, the gNB keeps removing expired ones, and the total number of reserved contexts does not grow. When the attack rate is high enough, the gNB reaches its capacity before any context expires. After that point, even if some contexts are released, the attacker continuously refills these spots with new incomplete connections, keeping the gNB in an overloaded state.
\section{Detection and Mitigation Approach}

\subsection{Detection Logic}

The main challenge in detecting an RRC signaling storm is to distinguish it from a legitimate high-load scenario. In both cases, the gNB may receive many connection requests and may become overloaded. However, in a high-load case, UEs that receive Msg4 normally continue the procedure and send Msg5, while an attacker does not.

For this reason, we follow the detection algorithm in~\cite{rrcstormdetection}, which relies on the ratio between the number of received Msg5 to the number of transmitted Msg4 during a sliding window:

\begin{equation}
R = \frac{N_{Msg5}}{N_{Msg4}}
\label{eq:ratio}
\end{equation}

 Under normal or high-load conditions, this ratio should remain close to one. During the attack, the ratio decreases because the attacker causes the gNB to send multiple Msg4 without completing the setup. Therefore, the gNB can detect the attack when this ratio drops below a configurable threshold during a sliding window.

\subsection{Pattern-Based Mitigation}

\begin{figure}[t]
    \centering
    \includegraphics[width=1.05\linewidth]{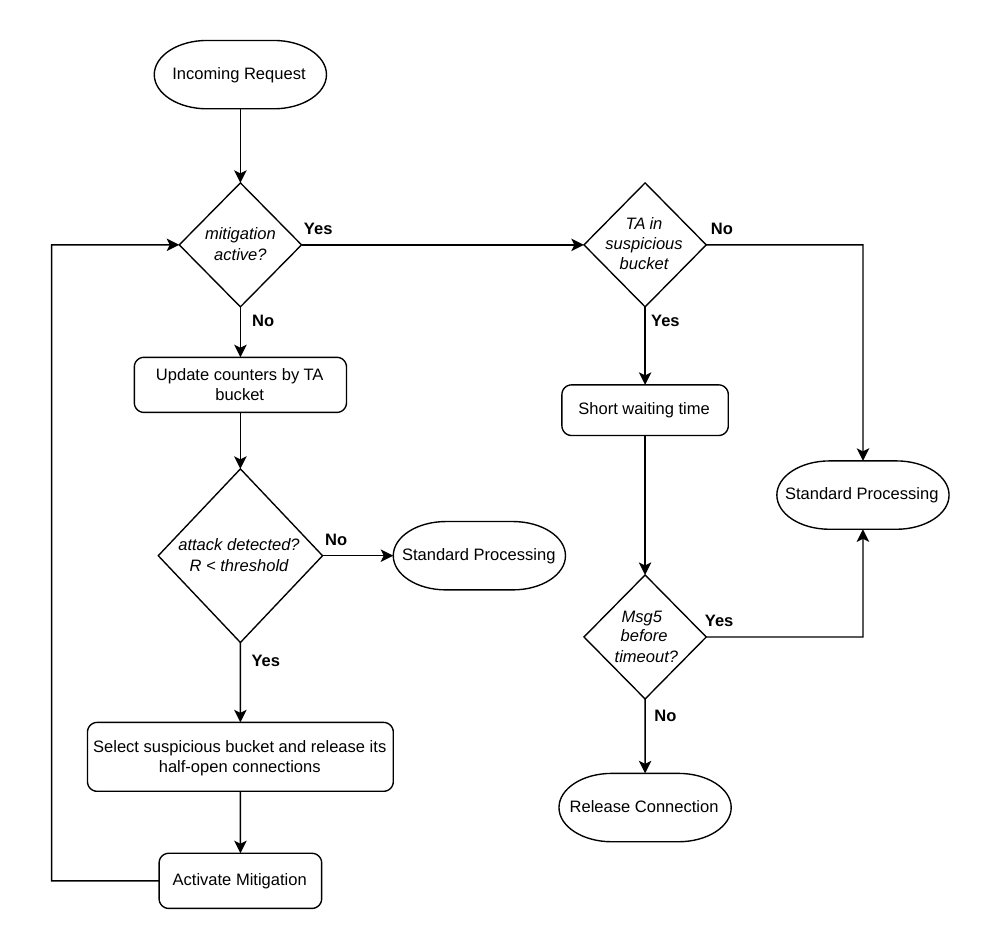}
    % \vspace{-0.5cm}
    \caption{Overview of the proposed pattern-based mitigation logic.}
    \label{fig:storm_mitigation}
    \vspace{-0.2in}
\end{figure}

After detecting the attack, the next step is to mitigate its impact. Fig.~\ref{fig:storm_mitigation} summarizes the proposed mitigation logic. Our mitigation technique follows a pattern-based approach. Instead of trying to identify the attacker directly, the gNB groups incomplete connection attempts according to observable features available during the random access phase. In our work, we use the TA value as the grouping feature.

The TA value is useful because it is estimated from the uplink signal received by the gNB and is not directly chosen by the UE. It provides an approximate indication of the UE distance from the gNB. The precision of the TA depends on the numerology. For example, with a subcarrier spacing of $30$ kHz, one TA step is approximately $39$ meters, while for $240$ kHz it becomes approximately $4.88$ meters. Thus, the TA is not precise enough to identify a device, but it can group requests that comes from a similar distance. The TA command value is standardized, ranging from $0$ up to a maximum of $3,846$ ~\cite{3gpp-38.211,3gpp-38.213}. %Note that we assume that the attacker does not change his TA value during the attack. 

The gNB maintains counters for half-open connections per TA bucket, where each bucket corresponds to a configurable range of TA values rather than necessarily one exact TA value. This is useful because, if the attacker is located near the boundary between two TA regions, its repeated RACH attempts may appear with two neighboring TA values. Grouping several TA values into one bucket makes the mitigation more robust to this small variation. When the attack is detected, the gNB selects the bucket with the largest number of half-open connections and marks it as suspicious. The gNB then releases the half-open contexts associated with this suspicious bucket, freeing reserved resources in the affected range immediately. 

For new connection attempts that fall into the same suspicious bucket while the attack is active, the gNB does not reject them immediately. Instead, it gives them a very short time window that is sufficient for a normal UE to complete the connection, but too short for an attacker to accumulate a large number of half-open connections.

This mitigation is intentionally lightweight, and legitimate UEs belonging to the same TA bucket as the attacker may be affected during the mitigation process, as they are required to complete the connection within a reduced time window with potential retries. However, this is a preventive measure that protects the rest of the cell from full resource exhaustion. The design is also extensible: the TA is only the first feature we use in our work, and future extensions can add additional features to build more accurate buckets.

It is important to note that our mitigation assumes the attacker stays roughly in one place. If the attacker moves during the attack, or uses several UEs at different locations, its connections would spread across multiple TA buckets instead of one, and no single bucket would clearly stand out as suspicious. In this case, our current mitigation would be less effective, since it only targets the bucket with the highest count of half-open connections. Detecting this kind of distributed attack would require adding features beyond TA, which we leave as future work.

\section{Testbed and Implementation}
\label{sec:testbed}

\begin{figure}
    \centering
    \includegraphics[width=\linewidth]{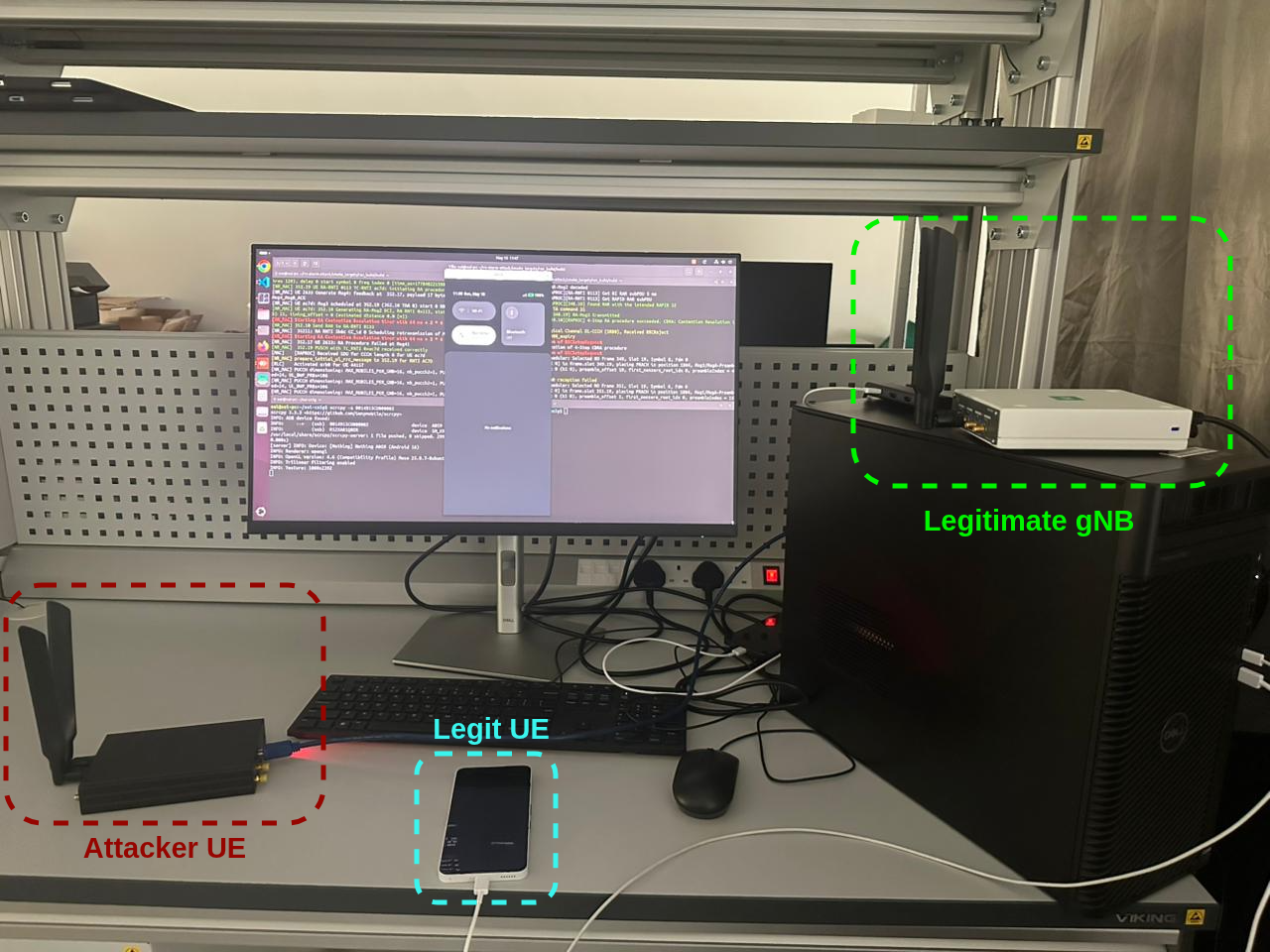}
    \caption{Experimental testbed used for the evaluation of the RRC storm attack and its mitigation technique.}
    \label{fig:storm-testbed}
    \vspace{-0.1in}
\end{figure}

\subsection{Experimental Testbed}
Our experiments were conducted on a real 5G testbed based on OAI. We selected OAI because it provides an open, widely used implementation of UE, gNB, and core network components~\cite{oai}. This was necessary for our work, since the attack requires changing the normal UE behavior during initial access, and the mitigation requires modifications to the gNB logic. 

Fig.~\ref{fig:storm-testbed} shows the experimental setup. The OAI system runs on a Dell Precision 3660 desktop running Ubuntu 24.04 LTS, and radio signals are transmitted using two USRP B210 software-defined radios~\cite{usrp-b210}. One SDR is used for the gNB, and the other for the UE.

Using the OAI UE instead of a commercial phone for the attacker is a practical requirement. Commercial UEs do not expose the low-level cellular logic needed to modify the RACH behavior, and such changes would require modem-level access. In contrast, OAI UE allows direct modification of the access procedure while remaining standards-based, making it suitable to prove that the protocol can be abused by a controllable UE.

To evaluate the impact on legitimate users, we also used a commercial phone - Nothing Phone (3a) - to test connection availability during the attack, with and without activation of our mitigation technique.

\subsection{Attack Implementation}

For the attack, we modified the OAI UE so that it does not send Msg5. Instead, it returns to an idle-like behavior and immediately starts a new RACH attempt with a newly generated random identity. This allows the attacker to continuously create incomplete connection attempts.

When the gNB becomes overloaded, it may respond with an RRCReject message. In normal operation, the UE then starts the T302 timer and refrains from initiating new connection attempts until its expiry~\cite{3gpp-38.331}. We disabled this behavior on the UE to ensure a continuous stream of requests that quickly occupies any newly available resources.

A practical detail is important here. Prior work describes the attack as repeatedly cycling through Msg1, Msg2, and Msg3 without waiting for Msg4~\cite{rrcstormdetection}. In our implementation, this was not sufficient. After Msg3, the attacker still had to receive Msg4 and acknowledge it at the lower layer. This behavior is consistent with the standard, which requires the UE to send a hybrid automatic repeat request (HARQ) acknowledgment upon successful reception of the contention resolution message~\cite{3gpp-38.321,3gpp-38.213}. Otherwise, the gNB quickly releases the temporary context, and half-open connections do not accumulate. In an emulated setup, restarting the RACH procedure from the RRC layer immediately after receiving Msg4 was sufficient. However, on the real radio testbed, this immediate restart was unreliable because the lower layers did not always have enough time to complete the Msg4 acknowledgment. Thus, we added a short timer after receiving Msg4 and before triggering the next RACH attempt. 
Therefore, in our implementation, the malicious UE proceeds up to Msg4, allows its acknowledgment to be transmitted, then skips Msg5 and starts a new RACH attempt after a short implementation timer.

\begin{figure*}
    \centering
    \includegraphics[width=\linewidth]{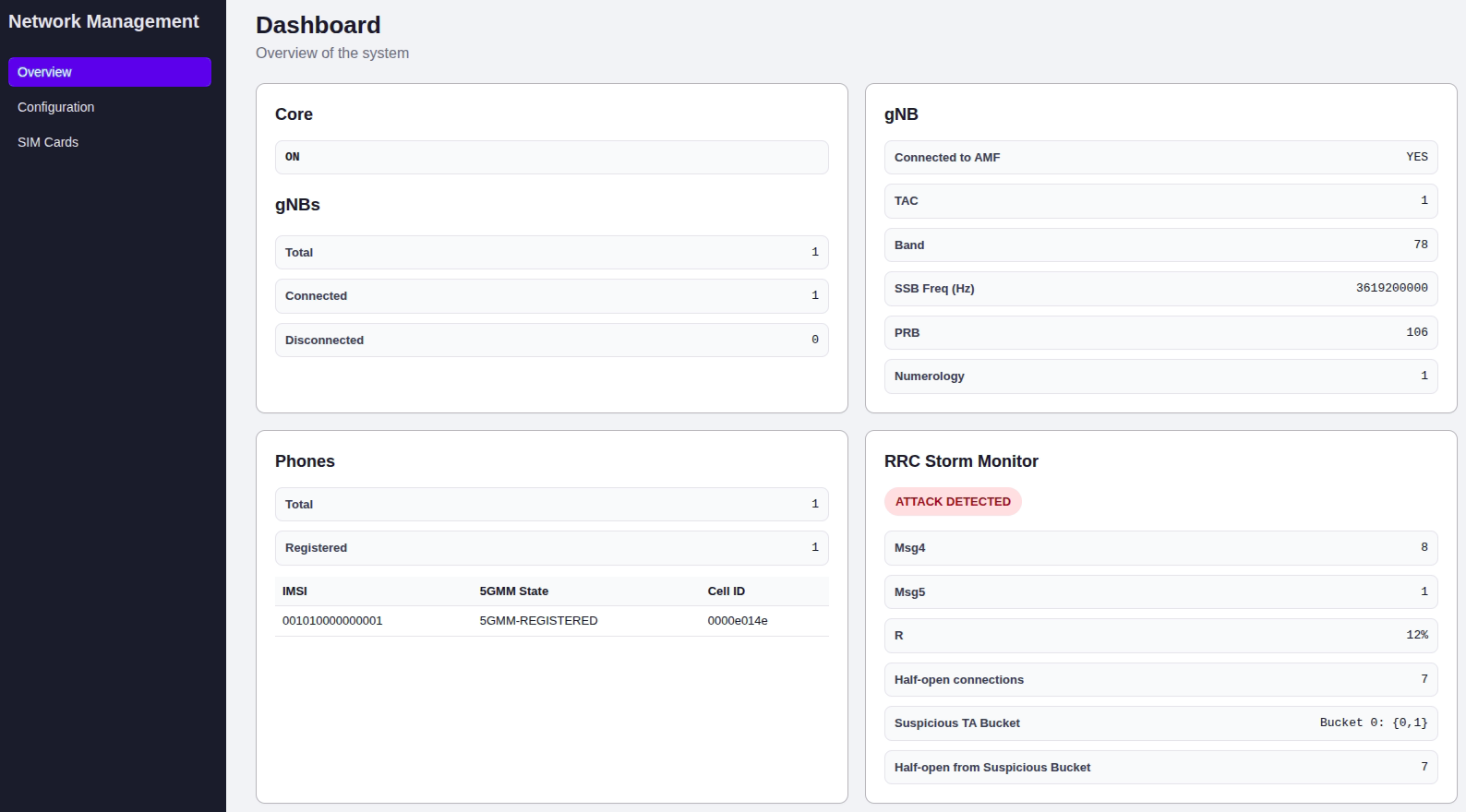}
    \caption{NMS interface with real-time system status and RRC storm indicators.}
    \label{fig:nms}
    \vspace{-0.2in}
\end{figure*}

\subsection{Detection and Mitigation Implementation}

Unlike prior work that deploys detection as an external application in an O-RAN architecture, our implementation operates directly inside the gNB. We integrate both detection and mitigation techniques within the RRC layer of the OAI gNB. This design choice allows us to access internal protocol features without relying on external interfaces and enables faster reaction to suspicious behavior. 

For detection, we implemented the monitoring logic inside the gNB RRC layer. The gNB counts Msg4 and Msg5 messages over a sliding time window and computes their ratio $R$ as in (\ref{eq:ratio}). The detection is triggered when this ratio falls below a configurable threshold, indicating a large number of incomplete connections. When the ratio rises above the threshold again, the system returns to normal operation.

Each time the gNB sends Msg4, the UE context is marked as half-open. The TA value obtained during random access is forwarded from the MAC layer to the RRC layer, where it is stored in the UE context. The gNB then increments the half-open counter associated with the matching TA bucket. The bucket size is configurable and can represent either a single TA value or a range of TA values.

When Msg5 is received, the context is marked as completed, and the half-open counter of the corresponding TA bucket is decremented. When an attack is detected, the mitigation selects the TA bucket with the highest number of half-open connections and marks it as suspicious. The half-open contexts associated with this suspicious bucket are then released immediately.

For new UEs falling within the suspicious bucket, the gNB starts a short timer after sending Msg4. If Msg5 is received before the timer expires, the UE is considered legitimate, and the timer is canceled. Otherwise, the UE context is released upon timer expiration.

\subsection{Network Management System}

To make the experiments observable and easier to reproduce, we developed a lightweight NMS.  It is a web-based interface used to configure and monitor the system while displaying relevant information from the core network, the gNB, and the attack status.

As shown in Fig.~\ref{fig:nms}, the NMS provides a real-time overview of the 5G system during normal operation. It displays the status of the core network, the gNB's availability, and whether it is successfully connected to the core network. It also provides information about connected UEs, including their international mobile subscriber identity (IMSI), registration state, and associated cell identifier. This allows us to observe how legitimate users interact with the network.

In addition to this general overview, the NMS includes a dedicated monitoring view for the RRC signaling storm. This view indicates whether the system is under attack and exposes the key metrics used by the detection and mitigation logic. In particular, it displays the computed detection ratio $R$ and the total number of half-open connections. It also highlights the identified suspicious TA bucket and the number of half-open connections associated with it.

The NMS is lightweight and containerized, enabling consistent deployment across different environments. It allows researchers to configure, execute, and monitor experiments systematically.

The full implementation of the NMS, along with the attack, detection, and mitigation mechanism, is publicly available~\cite{storm-github}.

\section{Evaluation}
\begin{figure}
    \centering
    \begin{subfigure}{0.48\linewidth}
        \centering
        \includegraphics[width=\linewidth]{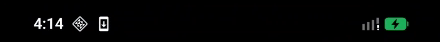}
        \caption{Without mitigation}
        \label{fig:phone_no_mitigation}
    \end{subfigure}
    \hfill
    \begin{subfigure}{0.48\linewidth}
        \centering
        \includegraphics[width=\linewidth]{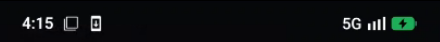}
        \caption{With mitigation}
        \label{fig:phone_with_mitigation}
    \end{subfigure}
    \caption{Real phone connectivity during the RRC storm attack.}
    \label{fig:phone_connectivity}

    \vspace{-0.2in}
\end{figure}
\subsection{Attack Evaluation}

We first evaluated the attack using the testbed described in Section~\ref{sec:testbed}. The goal is to verify whether the modified OAI UE can create enough incomplete RRC connections to exhaust the gNB resources and prevent legitimate users from accessing the cell.

In our experiments, the maximum number of supported UEs was configured in OAI. Although this value is configurable, it is still constrained by the practical capabilities of the setup and the gNB's resource allocation. In our testbed, the highest value that could be used reliably was $90$ UEs.  The waiting time for incomplete connections was configured to $5$ seconds.

We first evaluated the attack using the default OAI limit of $16$ UEs. After starting the malicious OAI UE, the number of half-open RRC connections increased quickly until the gNB reached its maximum capacity after approximately $0.65$ seconds. At that point, the gNB logs showed that no more UE contexts could be allocated, and new connection attempts were rejected using RRCReject messages.

We repeated the same experiment with the gNB configured to support $90$ UEs. In this case, the attack required more time to fill the available contexts, but the same behaviour was observed. The gNB reached its maximum configured capacity after approximately $3.61$ seconds.

Based on these two experiments, we estimate that the attacker generates incomplete connection attempts at an average rate of approximately $25$ requests per second. Therefore, if a gNB keeps incomplete contexts for a waiting time of $T$ seconds, an attacker operating at this rate can occupy approximately $25 \times T$ contexts before the first context is released, e.g., with a waiting time of $5$ seconds, the attacker would occupy around $125$ contexts before cleanup starts.

To confirm the practical impact on legitimate users, we also attempted to connect the Nothing phone 3a to the same gNB. Under normal conditions, the phone connected to the network successfully. During the attack, however, the phone failed to connect even after multiple attempts, as shown in Fig.~\ref{fig:phone_connectivity} (a). This confirms that the attack creates a denial of service (DoS) on commercial phones.

\subsection{Mitigation Technique Evaluation}

We then repeated the experiment with the mitigation mechanism enabled at the gNB. The same malicious OAI UE was used, and the attack was launched under the same testbed conditions.

During the attack, the gNB monitored the number of transmitted Msg4 messages and received Msg5 messages. As expected, the ratio $R$ decreased as the attacker continued creating incomplete connections. Once the ratio dropped below the configured threshold, set to $0.5$, the gNB detected the signaling storm and activated the proposed mitigation technique. More advanced and adaptive threshold selection strategies are discussed in~\cite{rrcstormdetection2}.

In our setup, each TA bucket was configured to cover two consecutive TA values. For example, the first bucket grouped TA values $0$ and $1$. During the attack, this bucket contained the largest number of half-open connections and was therefore selected as suspicious. This is expected because the attacker was physically close to the gNB in the testbed.
For new connection attempts coming from the suspicious TA bucket, the gNB started a short completion timer after sending Msg4. In our experiments, this timer was configured to $100$ ms, selected empirically based on the observed attacker rate (about $25$ requests per second, i.e., $40$ ms per request). As expected, this prevented the attacker from accumulating half-open contexts. Connection attempts from UEs outside the suspicious TA bucket were not affected by this short timer.

We also tested the effect of the mitigation on the Nothing phone 3a, whose TA value fell into the same suspicious bucket as the attacker's and was therefore subject to the same timer defined in the mitigation technique algorithm. The phone was able to connect to the network on either the first or the second attempt, as shown in Fig.~\ref{fig:phone_connectivity} (b). This shows the main trade-off of the mitigation: legitimate UEs in the suspicious bucket may experience stricter timing and occasional retries, but the cell remains available rather than being fully exhausted. Once connected, the phone reached the maximum throughput supported by our testbed. Finally, when the attack stopped, the ratio $R$ recovered. The gNB then ended the mitigation phase and returned to normal operation.

\section{Conclusion} 

In this paper, we revisited the RRC signaling storm attack in 5G systems from a practical perspective. We implemented the attack using OAI and evaluated it on a real testbed based on SDRs and commercial devices. The results showed that the attack can exhaust the capacity of the gNB and prevent real phones from connecting to the network. We also designed and implemented a lightweight gNB-side mitigation technique that monitors incomplete connections and applies a stricter connection completion timer to suspicious TA buckets. The mitigation limited the attacker's ability to exhaust gNB resources and preserved access for legitimate users under attack conditions.

By releasing the implementation as open source, we enable other researchers to reproduce the attack and extend the proposed defense. This can include improving the suspicious-bucket selection by combining TA with additional features, or by using machine learning-based techniques to more accurately distinguish attacker request patterns from real user behavior.

\bibliographystyle{IEEEtran}
\bibliography{sample-base} %references.bib

\end{document}